\documentclass[twocolumn,showpacs,preprintnumbers,amsmath,amssymb]{revtex4}

\usepackage{graphicx}
\usepackage{dcolumn}
\usepackage{bm}
\newcommand{\be}{\begin{equation}}
\newcommand{\ee}{\end{equation}}
\newcommand{\bea}{\begin{eqnarray}}
\newcommand{\eea}{\end{eqnarray}}
\newcommand{\ba}{\begin{array}}
\newcommand{\ea}{\end{array}}
\newcommand{\bi}{\begin{itemize}}
\newcommand{\ei}{\end{itemize}}
\newcommand{\lan}{\langle}
\newcommand{\ran}{\rangle}

\begin{document}

\title{Charmonium resonances and Fano line shapes}

\author{Xu Cao$^{1,2,3}$ }
\author{H. Lenske$^{2,4}${\footnote{Corresponding author: Horst.Lenske@theo.physik.uni-giessen.de}}}

\affiliation{$^1$Institute of Modern Physics, Chinese Academy of Sciences, Lanzhou 730000, China\\
$^2$Institut f\"{u}r Theoretische Physik, Universit\"{a}t Gie{\ss}en, D-35392 Gie{\ss}en, Germany\\
$^3$State Key Laboratory of Theoretical Physics, Institute of Theoretical Physics, Chinese
Academy of Sciences, Beijing 100190, China\\
$^4$GSI Darmstadt, D-64291 Darmstadt, Germany}
\date{\today}

\begin{abstract}

  Anomalous line shapes of quarkonia are explained naturally as an interference effect of a $c\bar c$ confined closed channel with the surrounding continua, well established in other fields of physics as Fano-resonances. We discuss a quark model coupled-channel analysis describing quarkonium as a mixing of closed $Q\bar Q$ and molecular-like $D\bar D$ open channels. The asymmetric line shapes observed in $\psi(3770)$ production cross sections in $e^+e^-$ annihilation to $D^0\bar{D}^0$ and $D^+ D^-$, respectively, are described very well. The method allows to extract directly from the data the amount of $Q\bar Q \leftrightarrow D\bar D$ configuration mixing.

\end{abstract}
\pacs {13.20.Gd, 13.25.Gv, 13.40.Gp, 13.66.Jn}
\maketitle{}


The observations of a multitude of unexpected states in the charmonium region is puzzling and a satisfactory explanation of their origin is pending since more than a decade. The aim of this letter is to point out the close resemblance of those sharp spectral structures to the phenomenon of bound states embedded into the continuum, $BSEC$ or $BIC$. In fact, long-lived states observed as sharp resonances above particle emission thresholds are an ubiquitous phenomenon in quantum systems of any scale.  The existence of such states was postulated already at the very beginning of modern quantum physics by von Neumann and Wigner \cite{Neum:1929}. In a seminal paper \cite{FanoPR1961} Fano revived that idea and established the effect in atomic physics explaining self-ionizing states as a quantum interference phenomenon of closed channel two-electron configurations with energetically degenerate single electron scattering states. Nowadays, modern laser techniques allow to simulate and manipulate such configuration mixing effects in atoms by a suitable choice of atto-second laser pulses \cite{Ott:2013}. BIC are observed abundantly in molecular spectra, see e.g \cite{Berb:1988,Ceder:2003}. Zhang et al. \cite{Zhang:2012} found the BIC phenomenon in the Hubbard model. Our own studies are showing that Fano-states do also exist in nuclei \cite{Orrigo:2005}, caused by the interference of closed channel multi-particle hole configurations with nucleon-core continuum states. A defining BIC-feature is the unusual small width-to-centroid energy ratio in the order of $10^{-3}$ or less.

In this letter we point out the relevance of the BIC-phenomenon for quarkonium physics. In the first place, our focus is on production cross sections, proposing a different view on quarkonium physics, adopting and exploring methods which are well established in other fields of physics. A cross-disciplinary Born-Oppenheimer approach to charmonium was also presented very recently by Braaten et al. \cite{Braaten:2014}. We keep the presentation of the formalism general and the results are equally well applicable to the heavy and the light quark sector. However, charmonium and bottonium states are of particular interest because of the appearance of sharp resonances with a width smaller by orders of magnitudes than the centroid mass. The steadily increasing number of unexplained so-called $XYZ$-states in the charmonium region indicates the need for extensions of the theoretical tool box. Coupling among open channels is a well studied problem in many areas of hadron physics, e.g. in meson production on the nucleon \cite{XuCao:2013}. The mixing of closed and open channels, however, is a less well established phenomenon in hadron physics.

\begin{figure}
  \begin{center}
{\includegraphics*[width=7.cm]{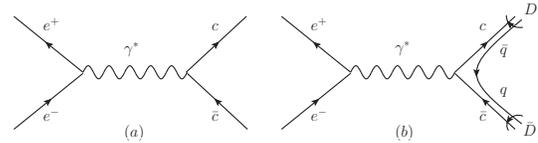}}
       \caption{
  Charmonium production by $e^+e^-$ annihilation: Population of the closed channel confined $c\bar c$ (left) and the open channel $D\bar D$ (right) components.
      \label{fig:Prod}}
  \end{center}
\end{figure}

The description of quarkonia including configuration mixing of confined $Q\bar Q$ heavy quark states and asymptotically emerging $D\bar D$ scattering states is in fact a demanding task. On the QCD level it is being treated by lattice calculations, e.g. \cite{LQCD2008}. A vast amount of work, however, is centered around the Cornell quark model \cite{Eichten1980,Eichten:2004,Barnes:2005b,Barnes:2005a,klshnkv:2005,Barnes:2008,Akleh:1996,Beveren:2009}, giving us a concrete physical model at hand for the spectroscopy of $Q\bar Q$ charmonium. Herein the $Q\bar Q$ spectra are obtained by a variational approach, using spherical harmonic oscillator wave functions as trial functions, see e.g. \cite{Barnes:2008,klshnkv:2005,Akleh:1996}. Results for a few bare charmonium states without continuum coupling are displayed in Tab. \ref{tab:Tab1}. Different from other approaches we use current quark masses \cite{pdg2010}. The potential parameters are $\alpha_S=0.55$, $\sigma=0.148$~GeV$^2$ for the Coulombic gluon exchange and the linear confinement potential, respectively, while for spin-singlet states we use $\sigma=0.088$~GeV$^2$. The shift constant is $C=0.014$~GeV. The quark model is used only as a spectroscopic tool for investigations of production reactions.

\begin{table}
\begin{center}
\caption{Charmonium spectrum of bare $c\bar c$ states without $D\bar D$ coupling obtained in the Cornell quark model compared to data \protect\cite{pdg2010}.\label{tab:Tab1}}
\begin{tabular}{|l| l| l| l|l|l|}
  \hline
  State         & M$_{c\bar c}$  & rms   & $\Gamma^{e^+e^-}_{c\bar c}$ & M$_{exp}$  & $\Gamma^{e^+e^-}_{exp}$ \\ \hline
  $n^{2S+1}L_J$ & [GeV] & [fm]  & [keV]         &  [GeV]        & [keV] \\ \hline
  $1^1S_0$      & 2.985 & 0.431 & $--$          & 2.981         & $--$ \\ \hline
  $1^3S_1$      & 3.098 & 0.391 & 5.266         & 3.096         & 5.55 $\pm$ 0.140  \\ \hline
  $2^3S_1$      & 3.676 & 0.860 & 1.879         & 3.686         & 2.35 $\pm$ 0.040\\ \hline
  $3^3S_1$      & 4.040 & 1.341 & 0.983         & 4.039         & 0.86 $\pm$ 0.070\\ \hline
  $1^3D_1$      & 3.819 & 0.887 & 7.917$\cdot 10^{-3}$  & 3.770 & 2.62 $\pm$ 0.018 \\ \hline
  $2^3D_1$      & 4.187 & 1.250 & 7.070$\cdot 10^{-3}$  & 4.153 & 0.83 $\pm$ 0.070 \\
  \hline
\end{tabular}
\end{center}
\end{table}

\begin{figure}
  \begin{center}
{\includegraphics*[width=7.cm]{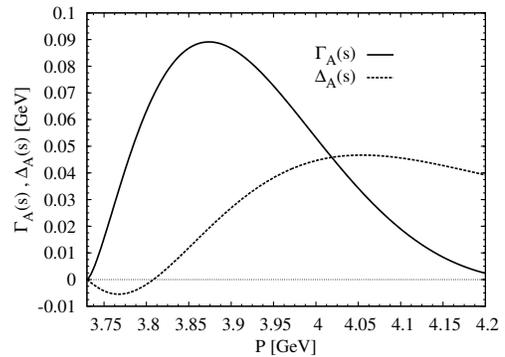}}
       \caption{
  Width $\Gamma_A$ and mass shift $\Delta_A$ for $\psi(3770)$ as obtained in the $^3P_0$ model for the intrinsic $c\bar c$ and $D=[c\bar q]$, $\bar D=[q\bar c]$ states using oscillator wave functions similar to \protect\cite{Barnes:2008,Akleh:1996,klshnkv:2005}.
      \label{fig:WidthShift}}
  \end{center}
\end{figure}

The interacting system is described in terms of the $Q\bar Q$ and the $D\bar D$ configurations as effective degrees of freedom. For the $D\bar D$ channel states the light quarks ($u,d,s$) are serving mainly to allow separation and rearrangement of the heavy $Q$ and $\bar Q$ quarks, respectively, by providing color neutrality through the formation of confined open charm states, as shown in Fig.~\ref{fig:Prod}. Rather than describing $D\bar D$ propagation as a four body problem we approximate the motion by an effective two-body meson configuration with the light quarks as spectators. This allows to reformulate the problem in terms of $Q\bar Q$ closed channels coupled to $Q\bar Q$ open channels which, however, are dressed by light quarks. The intrinsic D-meson structure is fully taken into account in the evaluation of matrix elements. The essential features of the approach are most easily accessed in terms of wave functions but obviously any other representation can be used equally well. With standard techniques \cite{BbS:1966} we reduce the coupled channels problem to a solvable problem in three-dimensional momentum space. In a partial wave representation the state vector is
\be
\Psi=\sum_A{z_A|A;P,q\ran}+\sum_{BC}{\int{dk k^2 z_{BC}(k)|BC;P,k\ran}}
\ee
given in terms of quarkonium states $|A;P,q\ran$ and $D\bar D$ channel states $|BC;P,k\ran$. The total four-momentum is $P^2=(p_B+p_C)^2=s$, where $s$ is the Mandelstam center-of-mass ($c.m.$) energy. Relative motion in the $Q\bar Q$  and the $D\bar D$ is described by the momenta $q$ and $k$, respectively. Since the kinetic energies are small compared to the rest masses we are able to use a separation \emph{ansatz},
\be
|A;P,q  \ran\simeq \varphi_P|A,q\ran \quad; \quad |BC;P,k\ran\simeq \varphi_P|BC,k\ran
\ee
decoupling the intrinsic degrees of freedom from the c.m. motion, given by the plane wave $\varphi_P$. $|A;q\ran$ denotes the quarkonium wave function in the rest frame. The $D\bar D$ scattering states are described by $|BC;k\ran$. The quarkonium states and the $D\bar D$ open channel states are interacting through an interaction $V$, taken from the $^3P_0$-model Lagrangian $\mathcal{L}_{int}=gm_q\bar{\Psi}_q\Psi_{\bar q}+h.c.$ where $\Psi_q$ denote quark Dirac field operators \cite{Micu:1969,Barnes:2008}. For simplicity, we consider the coupling of a single closed, confined $Q\bar Q$ channel to a set of open $D\bar D$ channels. The configuration coefficients are determined by the linear system
\bea
\left(s-s_A\right)z_A-2\sqrt{s_A}\sum_{BC}{\int^\infty_0{dkk^2 V_{A,BC}(k)z_{BC}}}=0 \label{eq:closed}\\
\left(s-s_{BC}\right)z_{BC}-2\sqrt{s_{BC}} V_{BC,A}z_A=0 \label{eq:open}
\eea
where $V_{A,BC}(k)=\lan A,q|V_{^3P_0}|BC;k\ran$ denotes the invariant matrix element of the $^3P_0$-interaction. The equations have to be solved as a partial wave boundary condition problem. How to construct such a solution is found e.g. in Fano's original paper \cite{FanoPR1961}. The essential, but not surprising result is that the channel coupling induces dispersive self-energies. Quarkonium states, for example, gain the dynamical self-energy $\Sigma_A(s)=\Delta_A(s)-\frac{i}{2}\Gamma_A(s)$ leading to a mass shift $\Delta_A$, a width $\Gamma_A$ and resulting in the finite life time $\tau_A\sim \frac{1}{\Gamma_A}$, where
\be
\Gamma_A(s)=2\pi\sum_{BC}{q_c\frac{E_B(q_c)E_C(q_c)}{\sqrt{s}}|V_{A,BC}(q_c)|^2}
\ee
where $q^2_c=(s-(M_B+M_C)^2)(s-(M_B-M_C)^2)/4s$ is the on-shell relative momentum in channel $c=[BC]$. Our results for the width and the mass shift of $\psi(3770)$, obeying a subtracted dispersion relation, are displayed in Fig. \ref{fig:WidthShift}.

In the open channels the coupling introduces an additional configuration phase shift, given by
\be
\cot{\delta_c(s)}=\frac{-s+m^2_A(s)}{\sqrt{s}\Gamma_A(s)} \quad .
\ee
where $m^2_A(s)=s^2_A+2\sqrt{s_A}\Delta_A(s)$. The wave function, normalized to $\delta(\sqrt{s}-\sqrt{s'})$, is
\be
|\Psi\ran=\sqrt{\frac{2}{\pi}}\frac{1}{\sqrt{\Gamma_{A}}} \left(\sin{\delta_c}|\Phi\ran+\cos{\delta_c}|\chi\ran\right)
\ee
The $Q\bar Q$ channel is now dressed by (virtual) admixtures of $D\bar D$ components,
\be
|\Phi\ran=|A;q\ran + \sum_{BC}{P\int^\infty_0dk k^2\frac{2\sqrt{s_k}V_{A,BC}(k)}{s-s_k}|BC;k\ran}
\ee
while the open channel component is given by
\be
|\chi\ran=\sum_{BC}{\frac{q_c}{\sqrt{s}}\frac{E_BE_C}{\sqrt{s}}V_{A,BC}(q_c)|BC;q_c\ran} \quad .
\ee
If there are n open channels, there are additional $j=1\cdots n-1$ solutions with $z^{j}_A=0$ given by superpositions of the open channel states only, contributing an incoherent smooth background \cite{FanoPR1961}. In atomic physics, those states are well known as \emph{dark states} \cite{Lambro:2007}. Here, we concentrate on the so-called \emph{bright} channel, including the resonance.

As an example we investigate the production of $\psi(3770)$ in leptonic $|\alpha\ran=|e^+ e^-\ran$ annihilation reactions, well studied experimentally~\cite{BESDDbar,BelleDDbar,CLEODDbar,BEShadrons,KEDRhadrons} and analysed by various theoretical approaches~\cite{HBLi2010,Zhang2010,YRLiu2010,Achasov2012,Chen2013}. Denoting the production operator by $\hat{T}$ the amplitude for the annihilation reaction $|\alpha\ran \to |\beta\rangle  = |\Psi\rangle$ is given by
\bea\label{eq:ProdAmp}
M_{\alpha\beta}=\lan\beta|\hat{T}|\alpha\ran=\sin{\delta_c}M^{(A)}_{\alpha\beta}(s)+\cos{\delta_c}M^{(BC)}_{\alpha\beta}(s)\\
M^{(A)}_{\alpha\beta}=\lan \Phi|\hat{T}|\alpha\ran \quad ; \quad M^{(BC)}_{\alpha\beta}=\lan \chi|\hat{T}|\alpha\ran \quad \label{Eq:ampM}
\eea
We define the (in general complex) Fano-line shape parameter $q_{\alpha\beta}=M^{(A)}_{\alpha\beta}/M^{(BC)}_{\alpha\beta}$ and the production cross section becomes
\be\label{eq:ProdSigma}
 \sigma_{\alpha\beta}=\sigma^{(BC)}_{\alpha\beta}\frac{|q_{\alpha\beta}-\cot{\delta_c}|^2}{1+\cot^2{\delta_c}}
\ee
The Fano-form factor, multiplying the open channel production cross section $\sigma^{(BC)}_{\alpha\beta}$, is, in fact, a scale-independent quantity: Irrespective of the scale of the physical system the same line shape will be observed for all systems with the same $q_{\alpha\beta}$, producing a line shape with an interference dip at energy $s=s_0$ where $cot\delta_c(s_0)=q_0$. By resolving such structures experimentally, important spectral information on the population of the (dressed) confined relative to the open channel components will be derived. The dependence of the Fano-parameter depends on the producing reaction, in principle provides the possibility to disentangle reaction dynamics and configuration dynamics by producing the same state by different probes.

\begin{figure}
  \begin{center}
{\includegraphics*[width=7.cm]{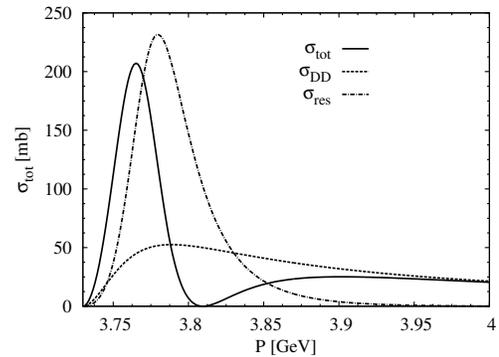}}
       \caption{
  Total cross section of $D\bar D$ elastic scattering, the partial contributions from the coupled channels resonance and the purely elastic $D\bar D$ scattering, produced by scalar and vector OBE potentials.
      \label{fig:partialDD}}
  \end{center}
\end{figure}

Eq.(\ref{eq:ProdSigma}) implies a separation of coupled channels effects from the intrinsic dynamics of the $Q\bar Q$ and the $D\bar D$ subsystems: $Q\bar Q$ and coupled channels dynamics are attached to the line shape parameter $q_{\alpha\beta}$ while $D\bar D$ dynamics is contained in $\sigma^{(BC)}_{\alpha\beta}$. An important conclusion is that the line shape seen in a production cross section must not be identical to the line shape observed in (hypothetical) $D\bar D$ elastic scattering. The latter would become visible in the $D\bar D$ partial wave total cross section, determined by non-resonant t-channel and resonant s-channel contributions. In order to avoid double-counting, $\sigma^{(BC)}_{\alpha\beta}$ must be evaluated without the coupling to those $A$-states which are treated explicitly. Here, we have to leave out the $\psi(3770)$ coupling. In order to estimate contributions of $D\bar D$ interactions we introduce scalar and vector one boson exchange (OBE) interactions where the coupling constants are chosen such that no $D\bar D$ bound states appear. The $D\bar D$ p-wave total cross sections with and without coupling to $\psi(3770)$ are shown in Fig. \ref{fig:partialDD}.

\begin{figure}
  \begin{center}
{\includegraphics*[width=4.2cm]{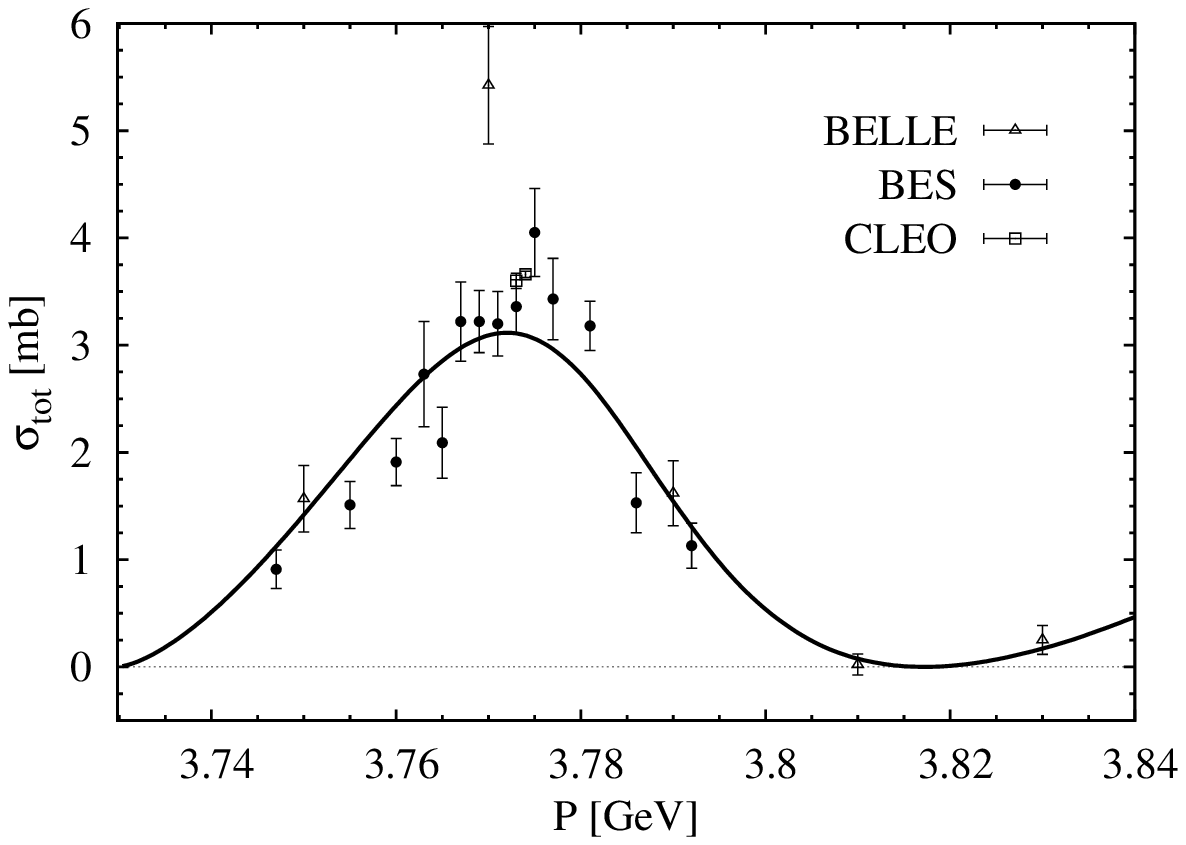}}
{\includegraphics*[width=4.2cm]{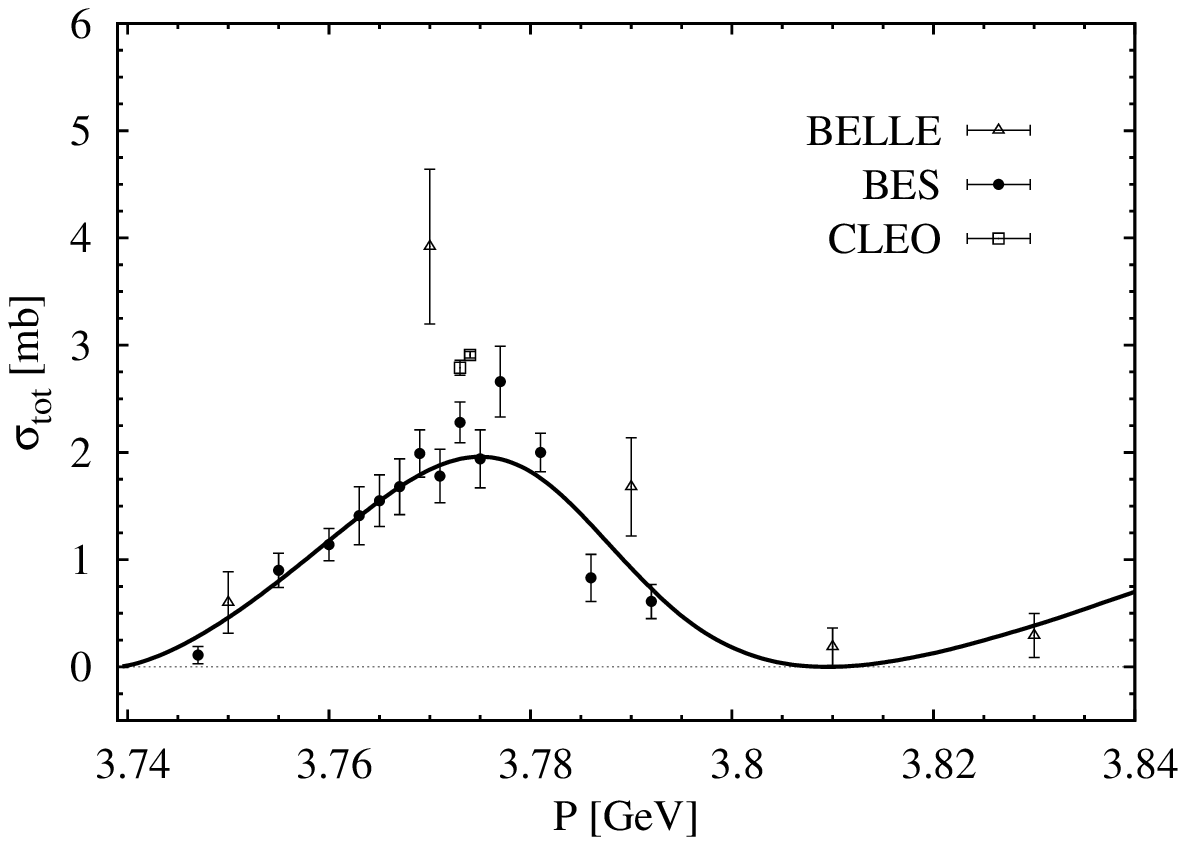}}
       \caption{
  Total cross sections of $e^+e^- \to D^0\bar{D}^0$ (right) and $e^+e^- \to D^+ D^-$ (left) reactions. Theoretical curves are compared to the data from \protect\cite{BelleDDbar,BESDDbar,CLEODDbar,BEShadrons}.
      \label{fig:fig5}}
  \end{center}
\end{figure}

With FSI the $D\bar D$ production amplitude is given by a Lippmann-Schwinger type equation accounting for the rescattering processes \cite{Achasov2012,Zhang:2012}
\be
M^{(BC)}_{\alpha\beta}=M^{(0)}_{\alpha\beta}+T^{(BC)}_\beta G^{(0)}_\beta M^{(BC)}_{\alpha\beta}
\ee
where $T^{(BC)}_\beta$ is the partial wave T-matrix of the uncoupled D-meson system and $M^{(0)}_{\alpha\beta}$ denotes the tree-level $e^+e^-\to D\bar D$ production amplitude. Within the quark model the direct $Q\bar Q$ production amplitudes are easily calculated \cite{Novikov:1978}. The ratio $\frac{\Gamma_{e^+e^-}(^3D_1)}{\Gamma_{e^+e^-}(^3S_1)}\simeq 50\left |\frac{R''_D(0)}{M^2_QR_S(0)}\right|^2$, indicates a strong suppression of the direct $^3D_1$ production as confirmed by Tab.\ref{tab:Tab1}. The observed widths, however, do not show that behaviour. Either the $^3D_1$ states contain an unexpected large amount of $^3S_1$ configurations or, according to Eq.(\ref{eq:ProdAmp}) they contain a substantial admixtures of $D\bar D$ components. The $D\bar D$ vertex, Fig. \ref{fig:Prod}, is not affected by that suppression. The $\gamma^*\to D\bar D$ vertex can be resolved into $\gamma^* \to Q\bar Q$ loops which subsequently decay into the final $D\bar D$ configuration. At the energies considered here, it is justified to replace the loops by a sum over the $^3S_1$, $^3D_1$ quark model states and the $D\bar D$ production will proceed mainly through $^3S_1$ intermediate states (or $^3S_1$ admixtures to $^3D_1$ states, respectively), as already pointed out by \cite{Achasov2012}. For $\psi(3770)$ this means that the $D\bar D$ component is mainly produced through the nearby $\psi'(3686)$ $^3S_1$ sub-threshold state. In Fig.~\ref{fig:fig5}, our final result for the production of $\psi(3770)$ are summarized. Treating for comparison the Fano-parameter as a free parameter leads to $q_0 = -0.4$ for $e^+e^- \to D^0\bar{D}^0$ and $q_0=-0.3$ for $D^+ D^-$ reactions, respectively. The different $q_0$ values in the neutral and charged $D\bar{D}$ channels are plausible considering the impact of $D^0 (D^+)$ mass gap and coulomb forces to the amplitudes in Eq.(\ref{Eq:ampM}). Both the line shapes of $\psi(3770)$ are convincingly well described.

In summary, we have proposed a novel view on charmonium line shapes observed in production cross sections by considering charmonium states as Fano resonances embedded into the $D\bar D$ continuum. The quark model was taken as a guidance to the internal structure of charmonium and open charm mesons. As an example we analysed the $e^+e^- \to D\bar{D}$ reaction populating the $\psi(3770)$ state. Quarkonium decay studies should take into account the coherent superposition of confined $Q\bar Q$ components and open charm (or bottom) continuum configuration providing a natural explanation for variations of line shapes in different production and decay channels. It is intriguing to apply the approach to other hadronic states, especially with respect of the newly found and hardly understood XYZ states.

\begin{acknowledgments}
This work was supported by the Deutsche Forschungsgemeinschaft ($DFG$) (CRC16, Grants No. B7), DFG project No. Le439/7 and in part by I3HP SPHERE, the LOEWE and the National Natural Science Foundation of China (Grant Nos. 11347146 and 11405222).
\end{acknowledgments}

\end{document}